\documentclass[prd,aps,showpacs,nofootinbib,eqsecnum,onecolumn,groupedaddress,amssymb]{revtex4}
\usepackage{amsmath}
\usepackage{amssymb}
\usepackage{amsfonts}
\usepackage{graphicx,bm}
\usepackage{dcolumn}
\usepackage{color,amsxtra}
\usepackage{epsf}
\usepackage{enumerate}
\usepackage{hhline}
\usepackage{array}
\usepackage{tabularx}
\usepackage{subfigure}
\usepackage{fancyhdr}
\usepackage{mathrsfs}
\newcommand{\be}{\begin{equation}}
\newcommand{\ee}{\end{equation}}
\newcommand{\bea}{\begin{eqnarray}}
\newcommand{\eea}{\end{eqnarray}}
\newcommand{\beaa}{\begin{eqnarray*}}
\newcommand{\eeaa}{\end{eqnarray*}}

\def\be{\begin{equation}}
\def\ee{\end{equation}}
\def\bea{\begin{eqnarray}}
\def\eea{\end{eqnarray}}

\begin{document}

\title{Mimetic covariant renormalizable gravity}
\author{
Ratbay~Myrzakulov$^{1}$
\footnote{E-mail address: rmyrzakulov@gmail.com},
Lorenzo~Sebastiani$^{1}$
\footnote{E-mail address: l.sebastiani@science.unitn.it}
Sunny~Vagnozzi$^{2, 3, 4, 5}$
\footnote{E-mail address: vagnozzi@nbi.dk}
and Sergio~Zerbini$^{6}$
\footnote{E-mail address: zerbini@science.unitn.it}
}
\affiliation{
$^1$Department of General \& Theoretical Physics and Eurasian Center for
Theoretical Physics, Eurasian National University, Astana 010008, Kazakhstan\\
$^2$ Niels Bohr International Academy and Discovery Center, Niels Bohr Institute, University of Copenhagen, Blegdamsvej 17, 2100 Copenhagen \O, Denmark \\
$^3$ Nordita,
KTH Royal Institute of Technology and Stockholm University,
Roslagstullsbacken 23, SE-106 91 Stockholm, Sweden \\
$^4$ The Oskar Klein Centre for Cosmoparticle Physics, Department of Physics, Stockholm University, AlbaNova, SE-106 91 Stockholm, Sweden \\
$^5$ ARC Centre of Excellence for Particle Physics at the Terascale, School of Physics, University of Melbourne, Victoria 3010, Australia \\
$^6$ Dipartimento di Fisica, Universit\`{a} di Trento, Italy, and TIFPA, Istituto Nazionale di Fisica Nucleare, Trento, Italy}

%\date{}

%%%%%%%%%%%%%%%%%%%%%
% Abstract
%%%%%%%%%%%%%%%%%%%%%
\begin{abstract}

Covariant renormalizable gravity is a Ho\u{r}ava-like extension of general relativity, enjoying full diffeomorphism invariance. However, the price to pay in order to maintain both covariance and renormalizability is the presence of an unknown fluid, whose non-standard coupling dynamically breaks Lorentz invariance. In this brief work we identify and explain the nature of this fluid, which we note describes the conformal mode of gravity, and arises naturally in frameworks such as that of mimetic gravity. Finally, we lay out extensions of the covariant Ho\u{r}ava-like model, which can serve as a guide for future model-building in this area. \\


\end{abstract}
%%%%%%%%%%%%%%%%%%%%%

%----------------------------
%\pacs{98.80.Cq, 12.60.-i, 04.50.Kd, 95.36.+x}
%\hspace{13.1cm}
%----------------------------

\maketitle

\def\thesection{\Roman{section}}
\def\theequation{\Roman{section}.\arabic{equation}}

Covariant renormalizable gravity (CRG hereafter) is an extension of Ho\u{r}ava gravity in which, unlike the latter, diffeomorphism invriance is preserved at the level of the action, only to be broken via a dynamical symmetry breaking. It in turn requires a new degree of freedom which might be thought of as playing a similar role to the Higgs field in the Standard Model. \\

Covariant renormalizable gravity was introduced in 2010 by Nojiri and Odintsov \cite{no}. Renormalizability is attained \textit{\`{a} la} Ho\u{r}ava \cite{gravity} by means of a modification to the graviton propagator, which in the UV is made to scale with spatial momenta $\textbf{k}$ as $1/\textbf{k} ^{2z}$. For a renormalizable theory of gravity in 3+1 dimensions, $z=3$ is required.\footnote{Here $z$ measures the anisotropy between space and time. Specifically, the theory is constructed so that it is compatible with anisotropic scaling with dynamical critical exponent $z$: $\textbf{x} \rightarrow b\textbf{x}, t \rightarrow b ^zt$.} However, whereas in \cite{gravity} such a modified behavior is obtained by the introduction of terms explicitly breaking diffeomorphism invariance, the model proposed in \cite{no} enjoys instead full diffeomorphism invariance. The breaking of Lorentz invariance which leads to the modified behavior of the propagator is established dynamically, via non-standard coupling to a perfect fluid, in a ``spontaneous symmetry breaking"-like fashion. The model was reformulated in \cite{proposal} by introducing an extra scalar field (whose gradient norm is constrained by a Lagrange multiplier in the action), the gradient of which plays the role of four-velocity of the unknown fluid. The action of the theory for $z=2n+2$ then takes the form:
\begin{eqnarray}
S = \int d^4x \sqrt{-g} \left [ \frac{R}{2\kappa ^2} - \alpha \left [ \left ( T ^{\mu \nu}\nabla _{\mu}\nabla _{\nu} + \gamma T \nabla ^{\rho}\nabla _{\rho} \right ) ^n \left (T ^{\mu \nu}R _{\mu \nu} + \beta TR \right ) ^2  \right ] - \lambda \left ( \frac{1}{2}\partial ^{\mu}\phi \partial _{\mu}\phi - U(\phi) \right ) \right ] \ ,
\label{action}
\end{eqnarray}
where $T _{\mu \nu} = \partial _{\mu}\phi \partial _{\nu}\phi$. The authors argue that in such a way it should be possible to bypass the appearance of unphysical modes, which presumably occur in \cite{gravity} due to the lack of full diffeomorphism invariance. Black hole and accelerating solutions of this model were studied in \cite{cognola}, while an extension within the $f(R)$ framework with very interesting cosmological applications had already been proposed prior to the scalar formulation, in \cite{odintsov}. \\

Nevertheless, in CRG full diffeomorphism invariance comes at a price, namely the presence of an exotic fluid, with equation of state parameter $w \neq -1,1/3$. The authors argue that the fluid might have an origin within the string theory framework. That is, it might correspond to heavy excited modes of strings. Here we note that \textit{this fluid can be readily interpreted as characterizing the conformal degree of freedom of gravity}. To see this, consider the version of the model formulated in \cite{proposal}, and whose action we have written explicitly above. The introduction of the additional scalar field is completely analogous to the appearance of an extra scalar degree of freedom in the recently proposed mimetic gravity \cite{mdm,cosmology} (see \cite{matarrese,matarrese2,matarrese3,matarrese4} for extensions of mimetic gravity and related frameworks, and further discussions especially on the role of disformal transformations). Here, the conformal mode of gravity is isolated in a covariant fashion, leading to the appearance of an additional scalar field whose gradient norm is fixed for the theory to be consistent (this constraint can be implemented through a Lagrange multiplier as well). It is shown that the dynamics of this additional scalar field can mimic cosmological dark matter. The scalar field appearing in mimetic gravity and that playing the role of velocity potential for the fluid in CRG are precisely the same field. \\

Mimetic gravity and CRG are equivalent in the limit where the coupling between the exotic fluid and gravity vanishes. That is, when $\alpha = 0$ in Eq.(\ref{action}), and $U _0 = -1/2$ to ensure the correct normalization for the gradient of the mimetic scalar field. Furthermore, mimetic gravity is also related to the scalar version of the Einstein-aether theory \cite{scalar}, which falls into a class of Lorentz-violating generally covariant extensions of General Relativity, where Lorentz invariance is dynamically broken by a unit timelike vector, the aether. Einstein-aether theories appear, under certain conditions, in the IR limit of projectable Ho\u{r}ava gravity \cite{aether}, where the lapse is a function of time alone. It is also a well-known fact that dark matter appears in the projectable version of Ho\u{r}ava gravity as an integration constant \cite{integration}. Not surprisingly, then, dark matter had already been identified in the CRG setup. In \cite{cognola}, the generalized Friedmann equations within this framework were obtained, and a term with energy density decaying with the scale factor $a$ as $a ^{-3}$ and magnitude fixed by an integration constant appeared. The term was subsequently discarded in order to study de Sitter solutions, but later ignored, while it is in fact mimicking the behavior of cosmological dark matter. In \cite{cognola} it was furthermore shown that CRG solutions include black hole, Einstein-space and accelerating solutions and satisfy the Area Law. This was done by working in the simplifying case where $\beta = 1$ (which it was argued corresponds to a minimal modification of General Relativity). \\

Having identified the nature of the unknown fluid in CRG, we now lay out extensions of this model which would be intriguing from the point of view of cosmology. On the one hand, one can add a suitable potential for the CRG scalar field which, analogously to the mimetic gravity framework, allows one to reconstruct basically any given expansion history of the Universe (see \cite{cosmology}).\footnote{Note that a model for unified inflation and late-time acceleration had already been proposed in the context of $f(R)$ CRG in \cite{odintsov}, motivated by the emergence of multiple de Sitter solutions. The cosmological structure of this theory is indeed very rich, featuring e.g. quintessence and phantom-type accelerations.} On the other hand, constructing the most general action for the CRG scalar field, by the addition of higher order derivative terms (see e.g. \cite{scalar}), can alter the sound speed and hence the dynamics of the dark matter on small scales, which could presumably solve the ``small-scale problems" of collisionless cold dark matter and the appearance of caustic singularities (see e.g. recent work carried out in \cite{mirzagholi}). At the same time, keeping the non-standard coupling between gravity and the fluid makes this extension of CRG an UV completion of the aforementioned models \cite{mirzagholi}, where the desirable properties of CRG are maintained, for instance the conjectured renormalizability. \\

Finally, it will be interesting to study static spherically symmetric solutions of generalized CRG (with the addition of a potential). This has been done in the case of mimetic gravity \cite{static}. In particular, with a suitable choice of potential it should be possible to reconstruct a solution which displays a linear correction to the usual well known Schwarszchild solution of general relativity. It has been argued that such linear correction might be used to explain the inferred flat rotation curves of spiral galaxies without the need for dark matter \cite{obri}. \\

In summary, we have explained the nature of the unknown fluid which is employed in covariant renormalizable gravity to break diffeomorphism invariance dynamically. The fluid, we argue, describes the conformal mode of gravity and has been identified in frameworks such as mimetic gravity. We lay out several extensions of CRG which can reproduce cosmologically interesting scenarios, as well as explaining phenomena on galactic scales.\footnote{Note that similar ideas have been discussed recently in \cite{inflfr}.} In an upcoming work we will study some of the proposed extensions, which in any case can serve as useful guide for future model-building in this area.

\section*{Acknowledgements}

We would like to thank Sergei Odintsov for useful discussions and comments on a draft of the manuscript. SV would like to thank the Niels Bohr International Academy for hospitality while this work was being completed.

%%%%%%%%%%%%%%%%%%%%%%%%%%%
%%%  Sec. I
%%%%%%%%%%%%%%%%%%%%%%%%%%%

\end{document}